\documentclass[10pt,a4paper,twoside]{article}

\usepackage{indentfirst}            
\usepackage{graphicx}               
\usepackage{amsgen,amsfonts,amssymb,amsbsy} 

\newenvironment{resum}{\begin{quote}\small}{\end{quote}}

\newcommand{\bfsf}[1]{\textsf{\textbf{#1}}}
\newcommand{\lf}[2]{\mbox{\Large $\frac{#1}{#2}$}}

\begin{document}

\thispagestyle{plain}       

\begin{center}


{\LARGE\bfsf{THE SPEED OF GRAVITY: WHAT A \\[15pt]  THEORY SAYS}}

\bigskip


\textbf{Antonio Alfonso-Faus }


\emph{E.U.I.T. Aeron\'autica,  Spain}\/

\end{center}

\medskip


\begin{resum}
Following a quantum-gravity approach we use a gravitational quantum defined elsewhere as well
as an effective gravitational ``cross section'' in conjunction with Mach's Principle and the
de Broglie wavelength concept. We find the speed of gravity $v_g$ related to a gravitational
mass m as given by $v_g \approx c (M_u/m)^{1/2}$, where c is the speed of light and $M_u$ the
mass of the seeable Universe. For the solar system we get $v_g \approx 10^{11} \,c$, which
agrees very well with experimental results. For the Universe $v_g \approx c$.

\end{resum}

\bigskip


\section{Introduction}

The concept of  ``action at a distance'' for gravity implies an infinite propagation speed. We
believe that there are no infinites of any kind in Nature: there is always a natural way to
avoid such infinites. Many arguments can be given in favor of a finite, but much larger than
the speed of light $c$, speed of gravity [1] including experimental results. If gravity is due
to some kind of interaction (as a quantum-gravity approach would demand) then the speed of
propagation of this interaction should be determined by a theory of gravity quanta [2]. We
know that there is no concept of force for gravity under a general relativity (GR) approach.
But GR is not a quantum theory. If one includes quanta for gravity then the propagation of the
interaction implying a force (a change of momentum with time by an absorption-emission
process) seems to be the appropriated approach. And then the question of the speed of this
propagation arises and a value much faster than light seems to be the case [1] on an
experimental basis.

Working with a gravity quantum and a gravitational cross section [2] we will find here the
speed of gravity using Mach's Principle and the de Broglie wavelength concept. It turns out
that the speed of gravity in general is many orders of magnitude much larger than the speed of
light (except for the Universe itself). To all intents and purposes this is practically an
infinite value. This result is in contrast with a relativity principle that considers the
speed of light as an absolute maximum. But relativity is consistent with faster-than-light
particles (tachyons). This means that our theory predicts a propagation of gravitational
interactions through tachyons in the form of gravity quanta.

\section{The Speed of Gravity
}

To determine the speed of gravity from a gravitational mass m one needs the concept of a
quantum of gravity of mass $m_g$ [2]:
\begin{equation}
  m_g = \lf{\hbar}{c^2 t}
\end{equation}
and a gravitational cross section $\sigma_g$  [2]:
\begin{equation}
  \sigma_g = \lf{Gm}{c^2}\cdot ct
\end{equation}
This gravitational cross section is the effective area of gravitational interaction for the
mass $m$. It is given by the product of its gravitational radius $Gm/c^2$ and the size of the
seeable Universe $ct$ ($t$ the age of the Universe). Interpreting Mach's Principle under the
view that the rest energy of any mass $mc^2$ is due to its gravitational potential with
respect to the mass of the rest of the seeable Universe $M_u$:
\begin{equation}\label{}
  \lf{GM_u m}{ct} = mc^2
\end{equation}
one has
\begin{equation}\label{}
  \lf{GM_u}{c^2} = c \, t
\end{equation}
Then the gravitational cross section of any mass given in (2) is the product of two
gravitational radiuses $Gm/c^2$ and $GM_u/c^2$.

The uncertainty principle for a quantum of gravity of mass $m_g$ and speed $v_g$ gives:
\begin{equation}\label{}
  m_g\,v_g\,\left(\sigma_g\right)^{1/2}\approx \hbar
\end{equation}
This is as much as to say that the de Broglie wavelength for the quantum of gravity has about
the size of the effective gravitational cross section:
\begin{equation}\label{}
  \left(\sigma_g\right)^{1/2}\approx \lf{\hbar}{m_gv_g}
\end{equation}
Then from (1), (2) and (5) we get
\begin{equation}\label{}
  v_g^2 \approx \lf{c^5}{G}\cdot\lf{t}{m}
\end{equation}
and using Mach's Principle in the form in (4) one finally has:
\begin{equation}\label{}
  v_g\approx c\,\sqrt{\lf{M_u}{m}}
\end{equation}

This is the speed of gravity as related to any mass $m$. For the Universe as a whole we have
$m = M_u$ and therefore $v_g\approx c$. There are about $10^{11}$ galaxies in the seeable
Universe. Then gravity from a galaxy has a speed of about $3\cdot10^5\, c$. There are about
$10^{11}$ equivalent solar systems in a galaxy. Then gravity inside a solar system has a speed
of about $10^{11}\,c$. Now, Van Flandern in [1] finds from experiments a lower limit for the
value of the speed of gravity $v_g \ge 2\cdot10^{10}\, c$, which implies a complete agreement
between our theory and the experiments.

\section{Conclusions}

Under a gravity quanta approach the speed of gravity $v_g$ is determined to be much faster
than light, $v_g \approx c \left(M_u/m\right)^{1/2}$. Then gravity propagates as tachyons,
consistent with the relativity principle. The experimental results do not contradict this
theoretical deduction.


\end{document}